\documentclass[conference]{IEEEtran}
\IEEEoverridecommandlockouts
\usepackage{graphicx}
\usepackage{cite}
\usepackage{amsmath}
\usepackage{amsfonts}
\usepackage{amssymb}
\usepackage{algorithm}
\usepackage{algorithmic}
\usepackage{xcolor}
\usepackage{psfrag}
\usepackage{epsfig}
\usepackage{multicol}
\usepackage{subcaption}
\usepackage{tikz}
\usepackage{subfloat}
\usetikzlibrary{fit}
\usetikzlibrary{positioning}
\usetikzlibrary{shapes,arrows}

\newlength\myindent
\newcommand{\qed}{\nobreak \ifvmode \relax \else
      \ifdim\lastskip<1.5em \hskip-\lastskip
      \hskip1.5em plus0em minus0.5em \fi \nobreak
      \vrule height0.75em width0.5em depth0.25em\fi}

\newcommand{\bm}{\mathbf}
\newcommand{\be}{\begin{equation}}
\newcommand{\ee}{\end{equation}}
\newcommand{\bse}{\begin{subequations}}
\newcommand{\ese}{\end{subequations}}
\newcommand{\bea}{\begin{eqnarray}}
\newcommand{\eea}{\end{eqnarray}}
\newcommand{\x}{{\bm x}}
\newcommand{\y}{{\bm y}}

\newcommand{\br}{{\bm r}}

\newcommand{\bA}{{\bm A}}

\newcommand{\bI}{{\bm I}}
\newcommand{\bR}{{\bm R}}

\newcommand{\bF}{{\bf F}}
\newcommand{\bD}{{\bf D}}

\newcommand{\bS}{{\bf S}}
\newcommand{\bG}{{\bf G}}
\newcommand{\bH}{{\bf H}}
\newcommand{\bX}{{\bf X}}

\newcommand{\bP}{{\bm P}}

\newcommand{\h}{{\bf h}}
\newcommand{\Ts}{{T_{\rm{s}}}}

\newcommand{\bh}{{\bf h}}

\newcommand{\bw}{{\bf w}}

\newcommand{\bs}{{\bf s}}
\newcommand{\bY}{{\bf Y}}
\newcommand{\by}{{\bf y}}

\newcommand{\bzero}{{\bf 0}}

\begin{document}

\raggedbottom

\title{Reduced Overhead Channel Estimation for OTFS With Split Pilot}
\author{\normalsize Danilo Lelin Li, Sanoopkumar P. S., Arman Farhang\\
Department of Electronic \& Electrical Engineering, Trinity College Dublin, Ireland \\
Email: \{lelinlid, pungayis, arman.farhang\}@tcd.ie.}

\maketitle

\begin{abstract}

Orthogonal time frequency space modulation (OTFS) is currently one of the most robust modulation techniques for high Doppler channels. However, to reap the benefits of OTFS, an accurate channel estimation is crucial. To this mean, the widely used embedded pilot structures use twice the channel length size as a delay guard to avoid interference between the pilot and data symbols.
Hence, incurring a large spectral efficiency loss, especially in wideband systems where the channel length is large. To reduce the pilot overhead, we propose a novel split pilot structure with two impulse pilots. With two pilots, we can use one to cancel the other, thus, capable of removing the pilot interference over data. To remove the data interference from the pilot, we also propose an iterative joint channel estimation and detection technique tailored to the proposed split pilot structure. With the interference caused by the delay spread solved, we reduce the number of delay guards in our system by half, significantly improving the spectral efficiency.
To corroborate our claims, we numerically demonstrate that our proposed method can achieve performance levels comparable to that of the full-guard method while using only half the delay guard. Additionally, we show that our proposed iterative channel estimating technique has a fast convergence speed, requiring only two iterations.

\end{abstract}

\vspace{0.15cm}

\vspace{-0.15cm}
\section{Introduction}

The 5th generation of wireless communication systems (5G) has rejoiced in the fast-paced development over recent years. However, the emerging applications and media-rich services such as autonomous driving, bullet trains and on board entertainment, in the 6th generation wireless systems (6G), more than ever require high levels of link reliability and extra low latency \cite{qos_5g,qoe_5g}. A common aspect of these applications is the highly dynamic and time-varying wireless environment due to mobility.
Retrieving the transmitted information signal impaired by such wireless channels is one of the main challenges in high mobility communication systems \cite{6g_channel}. 

Even though orthogonal frequency division multiplexing (OFDM) can easily handle frequency-selective channels, its performance significantly degrades in doubly-selective channels \cite{Hadani_wcnc_2017}. To address this issue, Orthogonal time frequency space (OTFS) modulation has emerged as an alternative candidate waveform for future networks that is robust to both time and frequency selective channels. OTFS deploys the delay Doppler (DD) domain for data transmission, where the wireless channel is sparse and its parameters vary at a much slower pace than in the time-frequency domain \cite{Hadani_wcnc_2017}.

The prevalent channel estimation technique in the present literature is a high-power impulse pilot embedded in the DD domain together with data symbols \cite{Embedded}. 
To avoid interference from the pilot to data symbols and vice versa, the pilot is surrounded by a number of zero-guards that are larger than the channel delay and Doppler spread. This leads to a high pilot overhead, specially as the system bandwidth increases and consequently, the discrete time baseband channel becomes longer.
To reduce the pilot overhead, the authors in \cite{Embedded} also presented a reduced guard embedded impulse pilot for integer Doppler scenarios by reducing the zero guard across the Doppler dimension to twice the maximum Doppler index. However, this technique is not suitable for practical scenarios with fractional Doppler. The authors in \cite{suerimposed_1,suerimposed_2,superimposed_3}
proposed a superimposed pilot structure to remove the pilot overhead completely. This technique superimposes a high-power pilot signal on the whole DD domain data. The main drawback of such a pilot structure is the huge computational complexity associated with the sophisticated estimation and detection algorithms. Furthermore, superimposed pilots are susceptible to channel variations within each time slot of OTFS, requiring a super-frame solely to obtain initial estimates of the channel delay and Doppler parameters. In \cite{Near_opt}, the authors developed a basis expansion model (BEM) based joint channel estimation and detection technique for impulse pilot with reduced guard across Doppler dimension. This method consists of deploying a reduced pilot overhead and iteratively refining the estimated channel and data using detected data as pseudo-pilots. However, using impulse pilots has the unique benefit of not requiring matrix inversion for channel estimation, a benefit that is lost in pseudo-pilots.

\begin{figure*}[hbt!]
\centering
\begin{subfigure}[b]{0.5\columnwidth}
	\centering
	\usetikzlibrary{fit}
\usetikzlibrary{positioning}
\usetikzlibrary{shapes,arrows}
 \usetikzlibrary{patterns} 
\usetikzlibrary{decorations.pathreplacing}

\begin{tikzpicture}[auto, >=triangle 45]
\def\a{-2};
\def\c{0};
\def\h{0.25};
\def\w{0.25};
\draw[fill=red!30](\a,\c)rectangle (\a+11*\w,\c+3*\h);
\draw[fill=red!30](\a,\c+8*\h)rectangle (\a+11*\w,\c+11*\h);
\draw[pattern color=red!30, opacity=0.3](\a,\c+6*\h)rectangle (\a+11*\w,\c+11*\h);

\draw[xstep=0.25 cm,ystep=0.25 cm,color=black] (\a,\c) grid (\a+11*\w,\c+11*\h);
\def\x{0};\def\y{5};
\draw(\a+\x*\w,\c+\y*\h)rectangle (\a+\x*\w+\w,\c+\y*\h+\h);
\node at (\a+\x*\w+0.125,\c+\y*\h+0.125){\tiny$ \triangle$};
\node at (\a+\x*\w+0.15,\c-0.25){{\footnotesize $n_{\text{p}}$}};
\node at (\a-0.25,\c+\y*\h+0.125){{\footnotesize $m_{\text{p}}$}};
\def\y{8};
\node at (\a+11*\w+0.125,\c+\y*\h+0.125)(a1){};
\draw[|<->|](a1)--node[rotate=90,xshift=-0.5 cm,yshift=-0.25 cm]{{\tiny $2L-1$}}++(0,-1.35);

\end{tikzpicture}
	\caption{Full-guard EP}
	\label{Fig:frame1}
\end{subfigure}
\hspace{-0.3 cm}
\begin{subfigure}[b]{0.6\columnwidth}
	\centering
	\usetikzlibrary{fit}
\usetikzlibrary{positioning}
\usetikzlibrary{shapes,arrows}

\begin{tikzpicture}[auto, >=triangle 45]
\def\a{-2};
\def\c{0};
\def\h{0.25};
\def\w{0.25};
\draw[fill=red!30](\a,\c)rectangle (\a+11*\w,\c+3*\h);
\draw[fill=red!30](\a,\c+6*\h)rectangle (\a+11*\w,\c+11*\h);
\draw[xstep=0.25 cm,ystep=0.25 cm,color=black] (\a,\c) grid (\a+11*\w,\c+11*\h);
\def\x{0};\def\y{5};
\draw(\a+\x*\w,\c+\y*\h)rectangle (\a+\x*\w+\w,\c+\y*\h+\h);
\node at (\a+\x*\w+0.125,\c+\y*\h+0.125){\tiny$ \triangle$};
\node at (\a+\x*\w+0.15,\c-0.25){{\footnotesize $n_{\text{p}}$}};
\node at (\a-0.25,\c+\y*\h+0.125){{\footnotesize $m_{\text{p}}$}};
\def\y{5};
\node at (\a+11*\w+0.125,\c+\y*\h+0.125)(a1){};
\draw[|-|](a1)--node[rotate=90,xshift=0.2 cm,yshift=-0.25 cm]{{\tiny $k$}}++(0,0.6);
\def\y{6};
\node at (\a+11*\w+0.125,\c+\y*\h+0.125)(a1){};
\draw[|<->|](a1)--node[rotate=90,xshift=-0.25 cm,yshift=-0.25 cm]{{\tiny $L$}}++(0,-0.85);
-
\end{tikzpicture}
	\caption{Reduced guard EP}
	\label{Fig:frame2}
\end{subfigure}
\hspace{-0.5 cm}
\begin{subfigure}[b]{0.5\columnwidth}
	\centering
	\usetikzlibrary{fit}
\usetikzlibrary{positioning}
\usetikzlibrary{shapes,arrows}

\begin{tikzpicture}[auto, >=triangle 45]
\def\a{-2};
\def\c{0};
\def\h{0.25};
\def\w{0.25};
\draw[fill=red!30](\a,\c)rectangle (\a+11*\w,\c+3*\h);
\draw[fill=red!30](\a,\c+6*\h)rectangle (\a+11*\w,\c+11*\h);
\draw[xstep=0.25 cm,ystep=0.25 cm,color=black] (\a,\c) grid (\a+11*\w,\c+11*\h);
\def\x{0};\def\y{5};
\draw(\a+\x*\w,\c+\y*\h)rectangle (\a+\x*\w+\w,\c+\y*\h+\h);
\node at (\a+\x*\w+0.125,\c+\y*\h+0.125){\tiny$ \triangle$};
\node at (\a+\x*\w+0.15,\c-0.25){{\footnotesize $n_{\text{p}}$}};
\node at (\a-0.25,\c+\y*\h+0.125){{\footnotesize $m_{\text{p}}$}};
\def\y{2};
\draw[fill=red!30](\a+\x*\w,\c+\y*\h)rectangle (\a+\x*\w+\w,\c+\y*\h+\h);
\node at (\a-0.55,\c+\y*\h+0.125){{\footnotesize $m_{\text{p}} +L$}};
\node at (\a+\x*\w+0.125,\c+\y*\h+0.125){$\times$};
\def\y{6};
\node at (\a+11*\w+0.125,\c+\y*\h+0.125)(a1){};
\draw[|<->|](a1)--node[rotate=90,xshift=-0.25 cm,yshift=-0.25 cm]{{\tiny $L$}}++(0,-0.85);
\draw(\a+14*\w,\c)rectangle (\a+14*\w+0.25,\c+0.25)node[xshift=0.65 cm,yshift=-0.15 cm,align=center]{{\small Pilot 2}};
\node at (\a+14*\w+0.125,\c+0.125){$\times$};
\def\c{0.5};
\draw(\a+14*\w,\c)rectangle (\a+14*\w+0.25,\c+0.25)node[xshift=0.65 cm,yshift=-0.15 cm]{{\small Pilot 1}};
\node at (\a+14*\w+0.125,\c+0.125){\tiny$ \triangle$};
\def\c{1};
\draw[fill=red!30](\a+14*\w,\c)rectangle (\a+14*\w+0.25,\c+0.25)node[xshift=0.5 cm,yshift=-0.15 cm]{{\small Data}};
\def\c{1.5};
\draw(\a+14*\w,\c)rectangle (\a+14*\w+0.25,\c+0.25)node[xshift=0.8 cm,yshift=-0.15 cm]{{\small Zero guard}};

\end{tikzpicture}
	\caption{Proposed split pilot}
	\label{Fig:frame3}
\end{subfigure}
\caption{Different pilot schemes}
\label{Fig:frame}

\end{figure*}
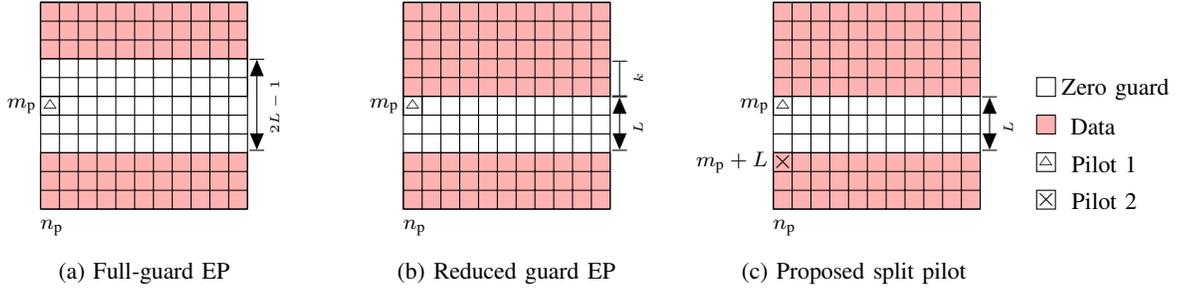

Multiple works in the literature have proposed to reduce the pilot overhead along the Doppler dimension. However, in practical scenarios, the channel length can be extremely large, and the delay guard becomes the main perpetrator of the large overhead. Furthermore, the channel length is directly proportional to the bandwidth of the system. Therefore, the required delay guard in wide band systems can become prohibitively large. To the best of our knowledge, no paper in the literature addressed reducing the pilot overhead along the delay dimension. Hence, we address the research question: 
\textit{``Can we reduce the delay guard of impulsive pilot schemes?"}

In this paper, we first introduce the challenges of reducing the delay guard from the embedded pilot (EP) structure. When the delay guard is reduced, the received pilot becomes contaminated with data symbols due to delay spread. Given that the pilot power is focused on a single impulse, it is still possible to perform channel estimation by absorbing the data interference as noise. However, the data information within the received pilot becomes irretrievable, incurring large performance loss. To solve this issue, we propose a split pilot structure with two impulse pilots. We use these two pilots to effectively remove the pilot interference from the received signal, without losing the data information. With our proposed structure, the pilot overhead is reduced by almost $50 \%$. To remove the data interference from the channel estimates, we developed an iterative joint channel estimation and detection technique. Our proposed technique uses the estimated data symbols to refine the channel estimates at each iteration. Through simulations, we show that our proposed technique significantly improves the BER performance of the split pilot structure. Furthermore, it has a high convergence speed, as two iterations are sufficient even in very high Doppler scenarios.

{\it Notations}: Matrices, vectors, and scalar quantities are denoted by boldface uppercase, boldface lowercase, and normal
letters, respectively. $[\bA]_{m,n}$
represents the element on row $m$ and column $n$ of $\bA$. $\bI_M$ and $\textbf{0}_{M\times N}$ are the $M \times M$ identity and $M \times N$ zero matrices, respectively. $(\cdot)^{\rm H}$ and  $|| \cdot ||_{\rm{F}}$ indicate Hermitian operation and Frobenius norm,  respectively. The function vec$(\bA)$ forms a vector by concatenating each column of $\bA$. $\bF_M$ is the normalized $M$-point discrete Fourier transform (DFT) matrix with the elements $[\bF_M]_{m,n}$ = $\frac{1}{\sqrt{M}}e^{\frac{-j2\pi mn}{M}}$, for $m,~n = 0,..., M-1$. 
\section{System Model}
\begin{figure*}[t!]
\normalsize

   {
   \be\label{eq:Hbc}\tag{5}
        \bH_{\rm{eff},p}=\begin{pmatrix}
        \bH_{0} & \bH_{N-1} & \cdots & \bH_{1}\\
        \bH_{1} &  \bH_{0} & \ddots & \bH_{2}\\
        \vdots & \ddots & \ddots & \vdots \\
        \bH_{N-1} &\cdots & \cdots & \bH_{0}
        
    \end{pmatrix}, \ \rm{ with } \ \bH_{n} = \begin{pmatrix}
        \bh[nL+L-1]  &\cdots & \bh[nL] & 0&\cdots & 0\\
        0  &\cdots & \bh_n[nL+1] & \bh_n[nL] &\cdots & 0\\
        \vdots & \ddots &\vdots & \ddots &\ddots &\vdots\\
        0  &\cdots & \bh[nL+L-1] &\cdots & \cdots & \bh_n[nL] 
        
    \end{pmatrix}
   \ee
   \be\label{eq:Dbc}\tag{6}
        \bS_{\rm{bc}}=\begin{pmatrix}
        \bS_{0} & \bS_{N-1} & \cdots & \bS_{1}\\
        \bS_{1} &  \bS_{0} & \ddots & \bS_{2}\\
        \vdots & \ddots & \ddots & \vdots \\
        \bS_{N-1} &\cdots & \cdots & \bS_{0}
        
    \end{pmatrix}, \ \rm{ with } \ \bS_{n} = \begin{pmatrix}
        [X_{\rm{p}}]_{L-1,n} & [X_{\rm{p}}]_{L-2,n}&\cdots & [X_{\rm{p}}]_{0,n}\\
        [X_{\rm{p}}]_{L,n} & [X_{\rm{p}}]_{L-1,n} &\cdots & [X_{\rm{p}}]_{1,n}\\
        \vdots & \ddots &\ddots &\vdots\\
        [X_{\rm{p}}]_{2L-2,n} & \cdots & \cdots & [X_{\rm{p}}]_{L-1,n}
        
    \end{pmatrix}
   \ee
   }

\hrulefill

\vspace*{0pt}
\end{figure*}

We consider an OTFS system in which each OTFS frame consists of $M$ delay bins with the delay resolution $\Delta \tau = \Ts $ seconds and $N$ Doppler bins with the Doppler resolution $ \Delta  \nu = \frac{1}{MN \Ts}$, where $\Ts$ is the sampling period.
At the OTFS transmitter, the information bits are mapped onto the $Q$-ary quadrature amplitude modulation (QAM) constellation points where $Q$ is the alphabet size. Then, they are embedded in the DD domain along with the pilot.  
Let the $M\times N$ matrices $\bD$ and $\bP$ represent the DD domain data symbols and the pilot signal, respectively, then the final DD domain transmitted signal $\bX$ is obtained after multiplexing $\bD$ and $\bP$, i.e., $\bX=\bD+\bP$. 
As the first stage of OTFS modulation, the DD frame is translated to the delay-time domain through an $N$-point inverse DFT (IDFT) matrix across the rows of $\bX$ \cite{Modemstructure}. The signal is then vectorized through a parallel-to-serial converter. To avoid interference between frames, a cyclic prefix (CP) of length $M_{\rm cp}$ is appended to the start of each frame. $M_{\rm cp}$ is chosen to be larger than the channel length $L$. The vectorized form of the transmit signal, of length $M_{\rm{T}}=MN+M_{\rm{cp}}$  can be represented as $\bs = \bA_{\rm{cp}}(\bF^{\rm{H}}_{N} \otimes \bI_M)\x$, where $\x=\rm{vec}(\bX)$ and  $\bA_{\rm{cp}} = [\bG_{\rm{cp}}^{\rm T} , \bI_{MN} ]^{\rm T}$ is the CP addition matrix and the $M_{\rm{cp}} \times MN$ matrix $\bG_{\rm{cp}}$ includes the last $M_{\rm{cp}}$ rows of $\bI_{MN}$.

The signal is transmitted through a linear time-varying (LTV) channel. Let the vector $\br$ represent the discrete-time baseband signal at the output of LTV channel. This signal can be represented as
\be
    \br = \bH \x + \bw,
\ee
where $\bw  \sim \mathcal{CN}(0,\sigma_{w}^{2}  \bI_{M_{\rm{T}}})$ is the complex additive white Gaussian noise (AWGN) vector with variance $\sigma_{w}^{2}$, and the $M_{\rm{T}}\times M_{\rm{T}}$ matrix $\bH$ is the LTV channel in the delay-time domain. The elements of $\bH$ can be represented as
\be
    [\bH]_{m,n} = \sum^{\Gamma-1}_{p=0} \alpha_p e^{j2\pi \upsilon_{p} n  T_{\rm{s}}} \delta [m-n -\ell_{p}],
\ee
where $\Gamma$ is the number of propagation paths and $\delta [\cdot]$ represents the Kronecker delta function. $\ell_{p}$, $\upsilon_{p}$, and $\alpha_p$ represent the delay tap, Doppler shift, and path gain associated with the $p$-th path. In practical wideband systems, the sampling rate is large enough to approximate the path delays to the nearest integer multiples of the sampling period. Hence, we only consider integer delays for $\ell_{p}$. This does not apply to the Doppler shifts, and $\upsilon_{p}$ includes both fractional and integer parts.

After removing CP and taking the delay-time domain received signal to the DD domain, we obtain the $MN\times 1$ received DD domain signal
\begin{align}\label{eq:Heff}    
    \begin{split}
        \by&=\bH_{\rm{eff}} \x +\bw,
    \end{split}
\end{align}
where $\bH_{\rm{eff}}= (\bF_{N} \otimes \bI_M) \bR_{\rm{cp}} \bH \bA_{\rm{cp}}(\bF^{\rm{H}}_{N} \otimes \bI_M)$ is the $MN \times MN$ effective channel matrix in the DD domain, $\bR_{\rm{cp}}=[\bzero_{MN \times M_{\rm{cp}}},\bI_{MN}]$ is the CP removal matrix and $\bF_N \otimes \bI_M$ takes the delay-time domain signal to the DD domain. The $M\times N$  DD received signal matrix $\bY$ can be obtained by rearranging the vector $\by$.

\section{Embedded Impulse Pilot Structure}

In this section, we present the main challenges that our proposed technique addresses. First, we introduce the widely used embedded 
pilot-based OTFS channel estimation. This technique requires long zero-guards along delay and Doppler, leading to a large pilot overhead. To improve the spectral efficiency, we reduce the delay guard above the pilot delay bin. With a reduced delay guard, the delay spread will cause the tail of the data above the pilot to contaminate the received pilot. Our derivations show that this tail can not be retrieved for data detection. Hence, a large portion of the data energy is lost leading to performance loss.

\subsection{Channel Estimation with Impulse Pilot}

For an EP-based channel estimation, a single impulse is placed in the DD bin $(m_{\rm{p}},n_{\rm{p}})$, surrounded by zero guards to avoid interference from data symbols, as shown in Fig. \ref{Fig:frame} (a). 
Let $\bY_{\rm{p}}$ represent the received pilot region in the DD domain, i.e., rows $m_{\rm{p}}$ to $m_{\rm{p}}+L-1$ of $\bY$. Due to the delay spread, the received pilot region is affected by the transmitted symbols within the pilot region and $L-1$ delay bins above. We define the data, pilot and full transmitted symbols that affect the received pilot region as $\bD_{\rm{p}}$,  $\bP_{\rm{p}}$ and $\bX_{\rm{p}}$, representing rows $m_{\rm{p}}-L+1$ to $m_{\rm{p}}+L-1$ of $\bD$, $\bP$ and $\bX$, respectively. We term this structure as full-guard since $\bX_{\rm{p}}$ is fully allocated for pilot, i.e., $\bD_{\rm{p}}=\bzero_{2L-1\times N}$. 

The vectorized representation of the received pilot region is given as
\be\label{eq:ypilot}
    \by_{\rm{p}}=\bH_{\rm{eff},p}\x_{\rm{p}} + \bw_{\rm{p}},
\ee
where $\by_{\rm{p}} = \rm{vec}(\bY_{\rm{p}})$, $\x_{\rm{p}} = \rm{vec}(\bX_{\rm{p}})$, $\bw_{\rm{p}}$ represents the AWGN noise in the received pilot region, and $\bH_{\rm{eff},p}$, given in \eqref{eq:Hbc}, is the reduced effective channel matrix reflecting only the received pilot region. The $LN\times 1$ vector $\bh$ contains the non-zero elements in the pilot column of $\bH_{\rm{eff}}$. It is worth noting that $\bH_{\rm{eff},p}$ is a block circulant channel matrix formed by the Toeplitz submatrices $\bH_{n} \in \mathbb{C}^{L\times (2L-1)}$. Hence, the property $\bH_{\rm{eff},p}\x_{\rm{p}} = \bS_{\rm{bc}}\bh$ 
can be obtained, where $\bS_{\rm{bc}}$ is the block circulant transmitted symbols matrix given in \eqref{eq:Dbc}. Similar to $\bS_{\rm{bc}}$, $\bS_{\rm{d}}$ can be derived with only the data symbols in $\bD_{\rm{p}}$ and $\bS_{\rm{p}}$ with only the pilot symbols in $\bP_{\rm{p}}$, i.e., $\bS_{\rm{bc}}=\bS_{\rm{d}}+\bS_{\rm{p}}$. It is worth noting that in the full-guard structure, $\bS_{\rm{d}}=\bzero_{LN\times LN}$, and when $n_{\rm{p}}=0$, $\bS_{\rm{p}}=\sqrt{\gamma_{\rm{p}}}\bI_{LN}$. Hence, we have  \setcounter{equation}{6}
\be\label{eq:D_bc}
    \bS_{\rm{bc}} = \bS_{\rm{d}}+ \bS_{\rm{p}}= \sqrt{\gamma_{\rm{p}}}\bI_{LN}.
\ee
Therefore, the channel estimates are obtained as
\be
    \widehat{\bh}=\frac{1}{\sqrt{\gamma_{\rm{p}}}}\by_{\rm{p}}=\frac{1}{\sqrt{\gamma_{\rm{p}}}}(\bS_{\rm{bc}}\bh +\bw_{\rm{p}})= \bh +\frac{1}{\sqrt{\gamma_{\rm{p}}}}\bw_{\rm{p}}.
\ee

The channel estimates $\widehat{\bh}$ only provide a single column of $\bH_{\rm{eff}}$. Hence,  
we bring the channel estimates $\widehat{\bh}$ to the delay-time domain and perform spline interpolation to find an estimate of the full channel matrix $\widehat{\bH}_{\rm{eff}}$ \cite{spline}. Finally, the data symbols are detected using this channel estimate.

The above channel estimates can only be achieved when $\bX_{\rm{p}}$ contains only pilot symbols, i.e., $(2L-1)N$ slots are allocated for the pilot. To improve the spectral efficiency, we analyze the effects of reducing the delay guard of the system in the following subsection.

\subsection{Channel Estimation Using Impulse Pilot with Reduced Guard}

Aiming to reduce the pilot overhead, we reduce the number of zero guard bins along the delay dimension. In this new structure, we assign $2L-1-k$ delay bins for pilot overhead, where $k$ represents the number of delay bins assigned for data instead of zero guards, e.g., $k=L-1$ shown in Fig. \ref{Fig:frame} (b). In this scenario, the received pilot region $\bY_{\rm{p}}$ remains in the same location, with the distinction that $\bX_{\rm{p}}$ contains both data symbols and the pilot signal.
We separate our derivations into two stages. The first stage consists of obtaining an initial channel and data estimates. In the second stage, we iteratively refine the channel estimates using the detected data from previous iterations.

\textit{Stage 1:} With a reduced delay guard, the delay spread will cause the tail of the data symbols above the pilot to spill into the received pilot region. Hence, $\bS_{\rm{d}}$ now contains the effects of data interference. With the pilot matrix $\bS_{\rm{p}}$ unchanged, the received pilot region can be shown as 
\be \label{eq:ypasdataandpilot}
    \by_{\rm{p}}=(\bS_{\rm{p}}+\bS_{\rm{d}})\bh + \bw_{\rm{p}} =(\sqrt{\gamma_{\rm{p}}}\bI_{LN}+\bS_{\rm{d}})\bh + \bw_{\rm{p}}.
\ee
Hence, by absorbing the interference from data into noise, the estimated channel is given as 
\be\label{eq:redguardchest}
    \widehat{\bh}=\frac{1}{\sqrt{\gamma_{\rm{p}}}}\by_{\rm{p}}= \bh +\frac{1}{\sqrt{\gamma_{\rm{p}}}}(\bS_{\rm{d}}\bh+\bw_{\rm{p}}).
\ee

As it is evident from \eqref{eq:redguardchest}, the channel estimate is imperfect. When using linear-based detectors, the imperfect channel leads to inaccuracies in data detection. This error is severely enhanced by the high pilot power, affecting the detected data. Hence, the pilot signal must be removed from the received signal before data detection. 
However, removing the pilot with \eqref{eq:redguardchest} also removes the tail of the data symbols above the pilot, i.e., $\bS_{\rm{d}}$. Therefore, a significant portion of the data energy is lost, leading to a significant performance loss. Hence, in the following stage, we iteratively refine the channel estimates, removing the data interference at each iteration. This also allows a more accurate removal of the pilot effect from the received signal.

\textit{Stage 2: }
To refine the channel estimates, we iteratively remove the data interference using the estimated data symbols from the previous iterations. Hence, the channel estimate at iteration $n$ is obtained as
\begin{align}\label{eq:esthn}
    \begin{split}
         \widehat{\bh}^{(n)}=&\frac{1}{\sqrt{\gamma_{\rm{p}}}}(\by_{\rm{p}}- \widehat{\bS}^{(n-1)}_{\rm{d}}\widehat{\bh}^{(n-1)})\\
         =&\bh +\frac{1}{\sqrt{\gamma_{\rm{p}}}}\left( 
 (\bS_{\rm{d}}- \widehat{\bS}^{(n-1)}_{\rm{d}})\bh +\bw_{\rm{p}} \right) \\
  &-\frac{1}{{\gamma_{\rm{p}}}}\widehat{\bS}^{(n-1)}_{\rm{d}}(\bS_{\rm{d}}\bh+\bw_{\rm{p}}).
    \end{split}
\end{align}
To simplify the above expression, we absorb the term $-\frac{1}{{\gamma_{\rm{p}}}}\widehat{\bS}^{(n-1)}_{\rm{d}}(\bS_{\rm{d}}\bh+\bw_{\rm{p}})$ into noise as generally the pilot power is sufficiently large to make this term negligible. From \eqref{eq:esthn}, it can be concluded that the channel estimates are refined if $\widehat{\bS}^{(n)}_{\rm{d}}$ converges to $\bS_{\rm{d}}$. 

After the channel estimate is refined, we remove the pilot effect from the received signal using \eqref{eq:esthn}
\be\label{eq:iterativeimpulse}
    \by_{\rm{p}} - \bS_{\rm{p}} \widehat{\bh}^{(n)}=\widehat{\bS}^{(n-1)}_{\rm{d}}\bh.
\ee
This indicates that the original data information within the received pilot region is replaced with the estimated data in the previous iterations. Hence, the iterative method also fails to preserve the tail of the data $\bS_{\rm{d}}$. 

We conclude that the main issue of reducing the delay guard comes from losing data energy within the received pilot region. To solve this issue, we propose in the next section a new pilot structure. Our proposed structure reduces the delay guard and allows us to remove the pilot effect without data energy loss.

\section{Proposed Split Pilot Structure and Channel Estimation Technique}

In the previous section, we showed that reducing the pilot delay guard in EP schemes causes data loss in the received pilot region. More specifically, this issue occurs when we remove the pilot effect from the received signal before data detection. Since data interference is present in the channel estimate, removing the pilot effect also removes a portion of the data energy, resulting in performance loss. This issue persists even when the data interference is iteratively removed from the channel estimate. To solve this issue, we propose a new pilot structure that is capable of removing the pilot effect without losing the data information. To refine the channel estimates, we propose an iterative joint channel estimation and detection technique tailored to our pilot structure. Ultimately, our proposed structure significantly reduces the pilot overhead compared to EP-based structures.

In our proposed pilot structure, we split the impulse pilot into two impulses of equal power $\frac{\gamma_{\rm{p}}}{2}$. We maintain the position of pilot 1 at DD bin $(m_{\rm{p}},n_{\rm{p}})$ and place pilot 2 at DD bin $(m_{\rm{p}}+L,n_{\rm{p}})$. We place the two impulses $L$ delay taps apart, hence the pilots do not interfere with each other. 
To minimize the pilot overhead, the second pilot is superimposed over data, and both pilots share a reduced zero guard region, as shown in Fig. \ref{Fig:frame} (c).  
Our proposed method is divided into two stages. In the first stage, we perform channel estimation absorbing the data interference into noise. To remove the pilot effect from the received signal, we cancel out the effect of one impulse with the response of the second impulse. Stage two consists of our proposed joint channel estimation and detection technique. In this stage, we refine the channel estimate by iteratively removing the data interference in the received pilot region. 

\textit{Stage 1:} Let us analyze the effect of each impulse pilot separately. 
Similar to \eqref{eq:ypilot}, we define the region of the received impulse pilot 1 and 2 as $\bY_{\rm{p}_1}$ and $\bY_{\rm{p}_2}$, respectively. Each region extends from the delay bin of the impulse pilot to the $L-1$ delay bins below.
The pilot regions are also affected by data interference $\bS_{\rm{d}_1}$ and $\bS_{\rm{d}_2}$, containing data symbols within $L-1$ delay bins above and after the respective pilots' location. 
Since there is minimal variation in the channel within $L$ delay taps, we disregard, only in our derivations, the channel variations within the two pilots. Hence, at each pilot region, we obtain the channel estimate
\begin{align}\label{eq:esth2}
    \begin{split}
        \widehat{\bh}_1=\frac{\sqrt{2}}{\sqrt{\gamma_{\rm{p}}}} \by_{\rm{p}_1}= \bh +\frac{\sqrt{2}}{\sqrt{\gamma_{\rm{p}}}}(\bS_{\rm{d}_1}\bh+\bw_{\rm{p}_1}),\\
        \widehat{\bh}_2=\frac{\sqrt{2}}{\sqrt{\gamma_{\rm{p}}}} \by_{\rm{p}_2}= \bh +\frac{\sqrt{2}}{\sqrt{\gamma_{\rm{p}}}}(\bS_{\rm{d}_2}\bh+\bw_{\rm{p}_2}),
    \end{split}
\end{align}
where $\bw_{\rm{p}_1}$ and $\bw_{\rm{p}_2}$ represent the AWGN noise from received pilot 1 and 2, respectively. To reduce the noise and interference, the final channel estimate $\widehat{\bh}$ is obtained after averaging $\widehat{\bh}_1$ and $\widehat{\bh}_2$.

As explained in the previous section, the pilot effect has to be removed from the received signal before data detection.
Therefore, we use $\widehat{\bh}_1$ to remove the effect of the second pilot
\begin{align}\label{eq:diffy1y2}
    \begin{split}
        \by_{\rm{p}_2}-\by_{\rm{p}_1}&=(\bS_{\rm{d}_2}-\bS_{\rm{d}_1})\bh+\bw_{\rm{p}_2}-\bw_{\rm{p}_1}.
    \end{split} 
\end{align}
In this expression, the pilot is completely removed, and the data information from both pilot regions is preserved. 
Furthermore, due to the shared delay guard between the two impulse pilots, there is no overlap between the non-zero elements of  $\bS_{\rm{d}_1}$ and $\bS_{\rm{d}_2}$. 
This ensures that the data symbols in $\bS_{\rm{d}_1}$ and $\bS_{\rm{d}_2}$ are discernible in the detector. Finally, the negative sign of the data vector $\bS_{\rm{d}_1}$ in \eqref{eq:diffy1y2} can be absorbed in the channel elements when reconstructing $\widehat{\bH}_{\rm{eff}}$. The received signal and channel estimate are then fed to the detector to obtain an initial data estimate $\widehat{\bD}^{(0)}$.

In reality, the channel ages within the two impulse pilots. The channel estimated from the first pilot is slightly different from the channel estimated from the second pilot. This introduces interference in \eqref{eq:diffy1y2}. To tackle this issue, we remove the pilot effect from the received signal after interpolating the channel estimates at each iteration of the next stage.

\begin{subfigures}

\begin{figure}
\end{figure}

\begin{algorithm}[t]

\begin{algorithmic}[1]

\caption{Proposed split pilot technique.}
\label{alg:Propsedtechnique}

\STATE Obtain the initial estimate $\widehat{\bh}=\frac{\widehat{\bh}_1+\widehat{\bh}_2}{2}$ using \eqref{eq:esth2}. 
\STATE Interpolate  $\widehat{\bh}$ to generate $\widehat{\bH}_{\rm{eff}}$.
\STATE Remove the pilot from $\bY$ using \eqref{eq:diffy1y2}.
\STATE Feed $\widehat{\bH}_{\rm{eff}}$ and $\bY$ to the detector to obtain $\widehat{\bD}^{(0)}$.

\FOR{$n=1 $ to $n_{\rm{max}}$}
    \STATE Refine channel estimates using \eqref{eq:esth2refined}.
    \STATE Interpolate  $\widehat{\bh}=\frac{\widehat{\bh}^{(n)}_1+\widehat{\bh}^{(n)}_2}{2}$ to generate $\widehat{\bH}_{\rm{eff}}^{(n)}$. 
    \STATE Remove Pilot 1 from $\bY$ using interpolated $\widehat{\bh}_2^{(n)}$.
    \STATE Remove Pilot 2 from $\bY$ using interpolated $\widehat{\bh}_1^{(n)}$.
    \STATE Feed $\widehat{\bH}_{\rm{eff}}^{(n)}$ and $\bY$ to the detector to obtain $\widehat{\bD}^{(n)}$.
    \STATE \textbf{Stop} if the stopping criteria is met.
\ENDFOR

\end{algorithmic}
\end{algorithm}

\end{subfigures}

\textit{Stage 2:} With an initial data estimation provided from stage 1, we can iteratively remove the data interference in \eqref{eq:esth2} and  refine the channel estimates. At each iteration $n$, we use the detected data from the previous iteration $\widehat{\bD}^{(n-1)}$ to generate the matrices $\widehat{\bS}^{(n-1)}_{\rm{d}_1}$ and $\widehat{\bS}^{(n-1)}_{\rm{d}_2}$. We then use these data matrices to remove the data interference from each pilot region before performing channel estimation. This way, the channel estimate from each impulse pilot at iteration $n$ is obtained as
\begin{align}\label{eq:esth2refined}
    \begin{split}
        \widehat{\bh}_1^{(n)}= \bh +\frac{\sqrt{2}}{\sqrt{\gamma_{\rm{p}}}}\left( 
 (\bS_{\rm{d}_1}- \widehat{\bS}^{(n-1)}_{\rm{d}_1})\bh +\bw_{\rm{p}_1} \right),\\
        \widehat{\bh}_2^{(n)}= \bh +\frac{\sqrt{2}}{\sqrt{\gamma_{\rm{p}}}}\left((\bS_{\rm{d}_2}- \widehat{\bS}^{(n-1)}_{\rm{d}_2})\bh +\bw_{\rm{p}_2} \right).
    \end{split}
\end{align}
Once again, we reduce the effect of noise by averaging the above channel estimates.

In \textit{Stage 1}, we considered the approximation of a time-invariant channel within each time slot. However, in high Doppler scenarios, the channel ages between the two pilots causing an interference in \eqref{eq:diffy1y2}. Although the channel only ages slightly within $L$ samples, this variation is greatly enhanced by the pilot power, causing performance loss in the system. Additionally, we have shown in \eqref{eq:iterativeimpulse} that removing the pilot effect with its own estimate leads to data energy loss. Therefore, to avoid both issues, we first interpolate the channel estimates in \eqref{eq:esth2refined} and use $ \widehat{\bh}_2^{(n)}$ to remove pilot 1 from the received signal and vice versa. The received signal after pilot removal becomes
\begin{align}\label{eq:propiter}
\hspace{-0.1 cm}
    \begin{split}
    \by_{\rm{p}_1}-\bS_{\rm{p}}\widehat{\bh}_2^{(n)}=&(\bS_{\rm{d}_1}-\bS_{\rm{d}_2}+ \widehat{\bS}^{(n-1)}_{\rm{d}_2})\bh+\bw_{\rm{p}_1}-\bw_{\rm{p}_2},\\
    \by_{\rm{p}_2}-\bS_{\rm{p}}\widehat{\bh}_1^{(n)}=&(\bS_{\rm{d}_2}-\bS_{\rm{d}_1}+ \widehat{\bS}^{(n-1)}_{\rm{d}_1})\bh+\bw_{\rm{p}_2}-\bw_{\rm{p}_1}.
    \end{split}
\end{align}
The received signal and channel estimate are then fed to the detector to obtain the data estimates $\widehat{\bD}^{(n)}$. We repeat this process until one of the following stopping criteria is met:
\begin{itemize}
    \item The data symbols after hard decision in $\bS_{\rm{d}_1}^{(n)}$ and $\bS_{\rm{d}_2}^{(n)}$ does not change from the previous iteration, i.e., $\bS_{\rm{d}_1}^{(n)}=\bS_{\rm{d}_1}^{(n-1)}$ and $\bS_{\rm{d}_2}^{(n)}=\bS_{\rm{d}_2}^{(n-1)}$.
    \item The maximum number of iterations $n_{\rm{max}}$ is reached. 
\end{itemize}

Comparing \eqref{eq:propiter} with \eqref{eq:iterativeimpulse}, we observe that our proposed method effectively retains the data information in the pilot region. Furthermore, the data interference reduces at each iteration, as $\widehat{\bS}^{(n)}_{\rm{d}_1}$ and $\widehat{\bS}^{(n)}_{\rm{d}_2}$ converges to ${\bS}_{\rm{d}_1}$ and ${\bS}_{\rm{d}_2}$, respectively.
The proposed technique is summarized in Algorithm \ref{alg:Propsedtechnique}.

\section{Simulation Results}

\begin{figure}
    \centering
    \includegraphics[scale=0.47]{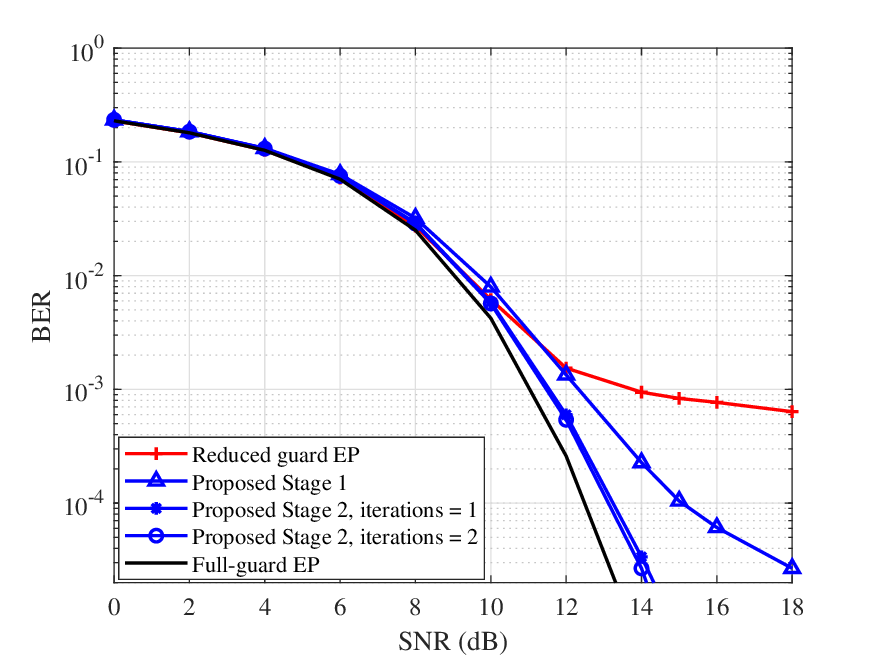}
    \caption{BER performance of our proposed scheme compared with full and reduced guard schemes.}

    \label{fig:BER}
\end{figure}

\begin{figure}
    \centering
    
    \includegraphics[scale=0.47]{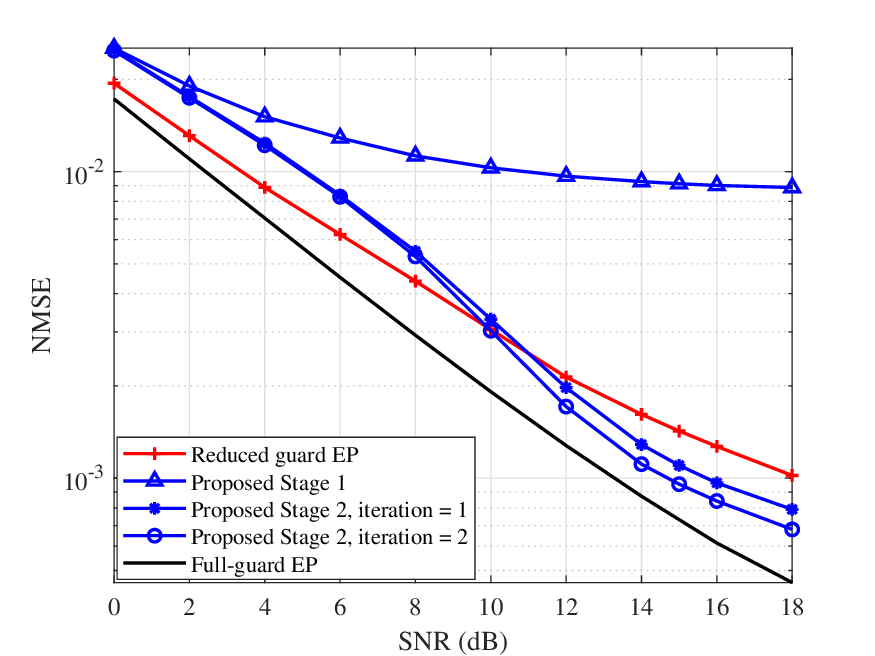}
    \caption{NMSE performance of our proposed scheme compared with full and reduced guard schemes.}
 
    \label{fig:NMSE}
\end{figure}

In this section, we confirm the efficacy of our proposed method comparing its performance with that of full and reduced guard schemes. We consider an OTFS system with $M=128$ delay and $N=32$ Doppler bins. In our
simulation setup, we use the carrier frequency $f_{\rm c}=5.9$~GHz,
subcarrier spacing $\Delta f = 15$~kHz, and 4-QAM modulation is deployed. 
The channel is modeled using the Extended Vehicular A (EVA) channel model, \cite{lte2009evolved}, and
the Doppler shift is generated using Jake’s model for the relative speed of $500$~km/h. The maximum delay spread of the channel is $2510$~ns and the sampling period of the system is $\Ts=520.3$~ns, leading to a channel length $L=5$. The full-guard scheme deploys $288$ DD bins as pilot overhead, while our proposed method deploys only $160$ DD bins for the pilot. We set the total pilot power at $\gamma_{\rm{p}}=40$~dB for all pilot schemes. We place Pilot 1 at delay bin $m_{\rm{p}}=\frac{M}{2}$ and Doppler bin $n_{\rm{p}}=0$. Otherwise stated, we consider four iterations for the EP with a reduced guard structure. We use the least squares minimum residual-based technique with interference cancellation \cite{lsmr} for data detection.

In Fig. \ref{fig:BER}, we compare the BER performance in function of data SNR of our proposed scheme with the full and reduced guard schemes. We show that the reduced guard scheme has a high BER floor even after refining the channel estimates for four iterations. This result is in line with the analysis in \eqref{eq:iterativeimpulse}, where we show that the iterative process is incapable of improving the data detection performance. In contrast, our proposed method can achieve similar performances after two iterations of refining the channel. The observed difference in performance is caused by the channel aging between the two pilots and noise enhancement within the pilot region, i.e., $\bw_{\rm{p}_2}-\bw_{\rm{p}_1}$ in \eqref{eq:diffy1y2}.

We also assess our proposed technique in terms of normalized mean square error (NMSE) of channel estimation, defined as $\frac{||\bH_{\rm{eff}}-\widehat{\bH}^{(n)}_{\rm{eff}}||_{\rm{F}}^2}{||\widehat{\bH}^{(n)}_{\rm{eff}}||_{\rm{F}}^2}$. Fig. \ref{fig:NMSE} shows that after convergence, our proposed technique reaches the NMSE of $10^{-3}$ with a 2~dB loss compared to EP with full-guard, whereas the reduced guard has over 4~dB loss. Additionally, it can be observed that even though the reduced guard EP attains acceptable NMSE at high SNRs, there is a BER floor (refer to Fig. \ref{fig:BER}) due to the data energy loss as shown in \eqref{eq:iterativeimpulse}.

\begin{figure}
    \centering
    \includegraphics[scale=0.47]{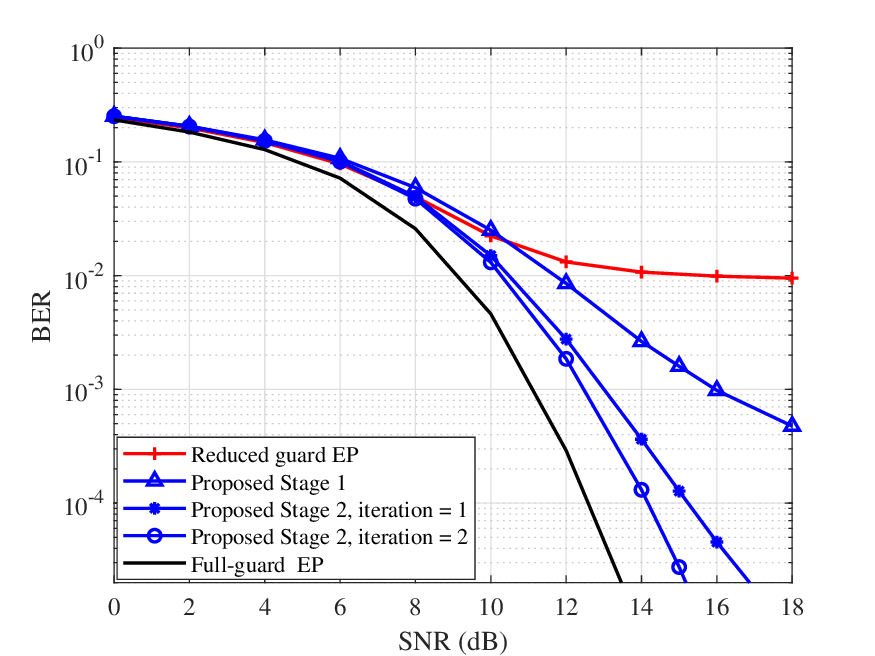}
    \caption{BER performance with increased channel length.}
 
    \label{fig:BERlargeL}
\end{figure}

We also show that our proposed method is still effective in scenarios with large channel lengths. In Fig. \ref{fig:BERlargeL}, we shorten the sampling period of the OTFS system to $\Ts = 133.33$~ns. Consequently, the channel length increases to $L=19$. In this system, the full-guard scheme deploys a total of 1184 DD bins for pilot overhead, while our proposed scheme requires only 608 DD bins. With a longer pilot region, the noise vectors in \eqref{eq:diffy1y2} increase. Consequently, the performance loss of our proposed method becomes slightly more pronounced due to the increased noise enhancement in the pilot region.

\begin{figure}
    \begin{subfigure}[]{0.45\columnwidth}
        \includegraphics[scale=0.47]{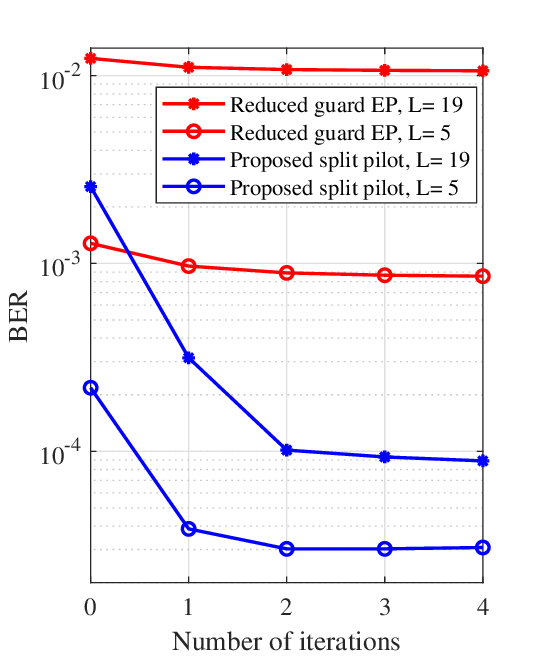}
        \caption{}
        \label{fig:ConvergenceBER}
    \end{subfigure}
    \hspace{0.1cm}
    \begin{subfigure}[]{0.45\columnwidth}
        \includegraphics[scale=0.47]{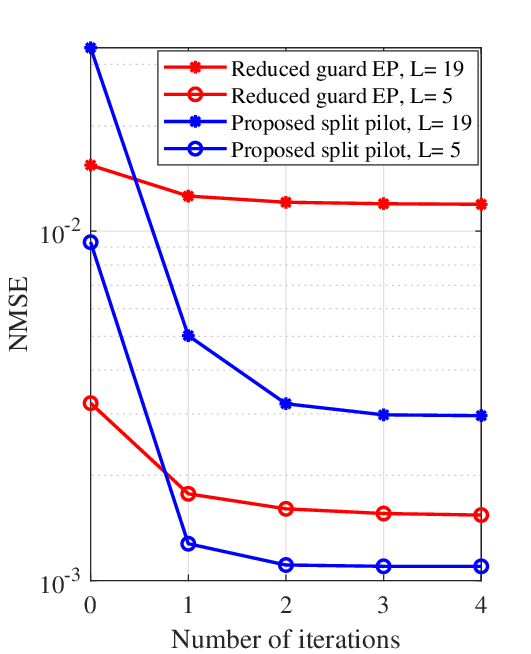}
        \caption{}
        \label{fig:ConvergenceNMSE}
    \end{subfigure}
    \caption{Convergence speed in terms of BER (a) and NMSE (b) performance at 14~dB.}
    \label{fig:Convergence}
\end{figure}

Finally, in Fig. \ref{fig:Convergence} we investigate the convergence speed of our proposed scheme in terms of BER and NMSE performance at SNR of $14$~dB. We show that within two iterations, our proposed method converges to significantly lower NMSE and BER values than the reduced guard EP. 

\section{Conclusion}

In this paper, we developed a new split pilot scheme that significantly reduces the pilot overhead in OTFS systems. Through a detailed theoretical analysis, we showed that reducing the delay guards in widely used EP degrades the estimation performance due to interference between data and pilots.  
As the pilot and data information become indistinguishable, all the data information in the pilot region is lost, defeating the purpose of reducing the pilot overhead. To overcome this issue, we propose a new scheme where the pilot is split into two. This allows for an effective way to remove the pilot interference over data information. The channel estimates can then be refined iteratively with the estimated data at each iteration. 
We show through simulations that our proposed method effectively saves almost $50 \%$ in pilot overhead and converges after two iterations even at high mobility scenarios.

\section{Acknowledgments}

This publication has emanated from research conducted with the financial support of Science Foundation Ireland under Grant numbers SFI/19/FFP/7005(T) and SFI/21/US/3757. For the purpose of Open Access, the authors have applied a CC BY public copyright license to any Author Accepted Manuscript version arising from this submission.

\bibliographystyle{IEEEtran}

\bibliography{main}

\end{document}